\documentclass[preprint,12pt,prb,amsmath,amssymb,floatfix]{revtex4}

\usepackage{bbold}
\usepackage{bm}
\usepackage{color}
\usepackage{dcolumn}
\usepackage{feynmf} 
\usepackage{graphics}
\usepackage{graphicx}
\usepackage{slashed}
\usepackage{placeins}
\usepackage{subfig}

\def\al{\alpha}

\def\veps{\varepsilon}

\def\be{\begin{equation}}
\def\ee{\end{equation}}
\def\bea{\begin{eqnarray}}
\def\eea{\end{eqnarray}}
\def\bse{\begin{subequations}}
\def\ese{\end{subequations}}
\def\bc{\begin{center}}
\def\ec{\end{center}}

\def\nonum{\nonumber}

\def\D{{\rm d}}

\newcommand{\comment}[1]{}

\catcode`,\active

\catcode`\,12

\begin{document}

\title{New results for a two-loop massless propagator-type \\ Feynman diagram}


\author{A.~V.~Kotikov$^1$ and S.~Teber$^{2,3}$}
\affiliation{
$^1$Bogoliubov Laboratory of Theoretical Physics, Joint Institute for Nuclear Research, 141980 Dubna, Russia.\\
$^2$Sorbonne Universit\'es, UPMC Universit\'e Paris 06, UMR 7589, LPTHE, F-75005, Paris, France.\\
$^3$CNRS, UMR 7589, LPTHE, F-75005, Paris, France.}

\date{\today}

\begin{abstract}
We consider the two-loop massless propagator-type Feynman diagram with an arbitrary (non-integer) index on the central line.
We analytically prove the equality of the two well-known results existing in the literature which express this diagram in terms of ${}_3F_2$-hypergeometric functions
of argument $-1$ and $1$, respectively. We also derive new representations for this diagram which may be of importance in practical calculations. 
\end{abstract}

\maketitle


\section{Introduction}

\begin{figure}[!tbp]
  \centering
  \subfloat[$J(\al_1,\al_2,\al_3,\al_4,\al_5)$]{\includegraphics{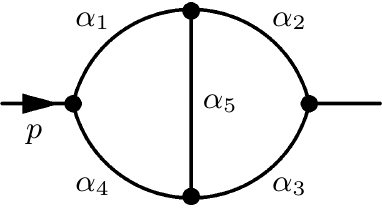}\label{fig:1}}
  \qquad \qquad \qquad \qquad
  \subfloat[$J(1,1,1,1,\al) \equiv J(\al)$]{\includegraphics{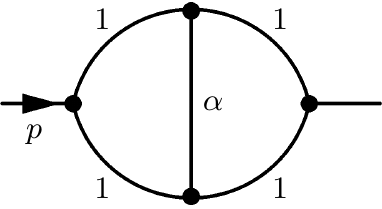}\label{fig:2}}
  \caption{Two-loop massless propagator-type Feynman diagrams.}
\end{figure}

The exact computation of the two-loop massless propagator-type Feynman diagram has been the subject of extensive studies 
over the last decades, see Ref.~[\onlinecite{Grozin:2012xi}] for a historical review.
The diagram is defined as:
\bea
J(\al_1,\al_2,\al_3,\al_4,\al_5) =
\int \int \frac{[\D^D k_1] \, [\D^D k_2]}{k_1^{2\al_1}\,k_2^{2\al_2}\,(k_2-p)^{2\al_3}\,(k_1-p)^{2\al_4}\,(k_2-k_1)^{2\al_5}} \, ,
\label{def:J}
\eea
where $[\D^D k] = \D^D k /(2\pi)^D$, the $\al_i$ (the five powers of the propagators) are the so-called indices of the diagram and $p$ is the external momentum in an Euclidean space-time of dimensionality $D$, see
Fig.~\ref{fig:1} for a graphical representation. When all five indices are integers, the diagram is easily computed, {\it e.g.}, with the help of
the integration by parts (IBP) procedure [\onlinecite{Vasiliev:1981dg,Chetyrkin:1981qh}]. On the other hand, for arbitrary (non-integer) indices an exact evaluation becomes highly non-trivial
and peculiar cases have to be considered, see,  {\it e.g.},
Refs.~[\onlinecite{Chetyrkin:1980pr,Vasiliev:1981dg,Chetyrkin:1981qh,Kazakov:1983ns,Kazakov:1983pk,Broadhurst86,Gracey:1992ew,KivelSV93,Kotikov:1995cw,Broadhurst:1996ur,Broadhurst:1996yc,Broadhurst:2002gb,Bierenbaum:2003ud,Kotikov:2013kcl,Kotikov:2013eha,Teber:2012de,Hathrell:1981zb}].
Among these peculiar cases, the simplest non-trivial diagram is the one with an arbitrary index on the central line:
\be
J(1,1,1,1,\al) =
\int \int \frac{[\D^D k_1] \, [\D^D k_2]}{k_1^{2}\,k_2^{2}\,(k_2-p)^{2}\,(k_1-p)^{2}\,(k_2-k_1)^{2\al}} \equiv J(\al)\, ,
\label{def:Jal}
\ee
see Fig.~\ref{fig:2}. The diagram (\ref{def:Jal}) has been extensively studied in the past, see the review [\onlinecite{Grozin:2012xi}], 
and its exact computation has led to a number of important applications over the years. 
As a first well known example, let us mention the analytical evaluation of the $5$-loop $\beta-$function of the $\Phi^4$ model, 
see Refs.~[\onlinecite{Kazakov:1983ns,Kazakov:1983pk}] and discussions and references therein. Secondly, 
the most complicated contributions to the values of critical exponents computed within a $1/N$-expansion, see
Refs.~[\onlinecite{Vasiliev:1981dg,KivelSV93}] as well as Vasil'ev's textbook [\onlinecite{VassilievBook}], depend, at order $1/N^2$, on the derivative with respect to the index of the central line of $J(\al)$; 
the fully analytic computation of the derivative has been obtained in Ref.~[\onlinecite{Broadhurst:1996yc}] 
based on the results of Ref.~[\onlinecite{Kotikov:1995cw}]. More recently, it was realized in Ref.~[\onlinecite{Teber:2012de}]
that general multiloop techniques such as those developed in Refs.~[\onlinecite{Vasiliev:1981dg,Chetyrkin:1981qh}] as well as the sophisticated
developments brought by Refs.~[\onlinecite{Kazakov:1983ns,Kazakov:1983pk,Kotikov:1995cw}] were of crucial importance for the exact computation of
interaction correction effects in brane world-like effective field theories describing some modern planar condensed matter physics 
systems, see also Refs.~[\onlinecite{Kotikov:2013kcl,Kotikov:2013eha,Teber:2014ita}] 
for developments in this direction as well as Ref.~[\onlinecite{Teber:2016unz}] and references therein for a short review on some of these aspects.  

\subsection{Existing results for $J(\al)$}

In what follows we shall use dimensional regularization and take $D=4-2\veps$. The dependence of
$J(\al)$ on momentum is trivial as it follows from simple dimensional analysis. We may then write Eq.~(\ref{def:Jal}) as
\be
J(\al) = \frac{p^{2 d_F}}{(4\pi)^D}\,I(\al)\, ,
\ee
where 
\be
d_F = D-4-\al= - \al -2 \veps\, ,
\ee
is the dimension of the diagram and $I(\al)$ its so-called (dimensionless)
coefficient function. It is the later that is of interest as it is in general non-trivial to compute. 
For most diagrams, the coefficient function is known only in the form of the first few terms of the Laurent series in $\veps$. In the case of $I(\al)$
exact results are available and we shall focus on them.

There are actually two different results which can be found in the literature for 
$I(\al)$.~\footnote{Two other results for this diagram can be found in Refs.~[\onlinecite{Hathrell:1981zb}] and [\onlinecite{Broadhurst:1996ur}], see the review [\onlinecite{Grozin:2012xi}] for a presentation of these results.}   
The first one, Ref.~[\onlinecite{Kazakov:1983pk}], has been derived by solving functional equations obtained with the help of a combination of the IBP
procedure [\onlinecite{Vasiliev:1981dg,Chetyrkin:1981qh}] together with several other transformations  [\onlinecite{Vasiliev:1981dg,Gorishnii:1984te,Broadhurst86}].
The result contains a one-fold series and reads:
\bea
I(1+\al) &=&
-2\, \frac{\Gamma^2(1-\veps)\Gamma(\veps)\Gamma(-\veps-\al) \Gamma(\al+2\veps)}{\Gamma(2-2\veps)} \Biggl[ 
\frac{1}{\Gamma(1+\al)\Gamma(1-3\veps-\al)}
\label{Kazakov}\\
&\times& 
\sum_{n=1}^{\infty}\,(-1)^n \frac{\Gamma(n+1-2\veps)}{\Gamma(n+\veps)}\,\left(\frac{1}{n+\al+\veps} + \frac{1}{n-\al-2\veps}\right)
+\cos [\pi \veps] \Biggr]\, .
\nonum
\eea
Notice that the above one-fold series can be represented as a combination of two ${}_3F_2$-hypergeometric functions of argument $-1$.

The second result, Ref.~[\onlinecite{Kotikov:1995cw}], is based on an application of the Gegenbauer polynomial technique that in-turn is a generalization of earlier studies
[\onlinecite{Chetyrkin:1980pr}].  It expresses the function $I(1+\al)$ in terms of a single ${}_3F_2$ function of argument $1$ with the result reading:
\bea
I(1+\al) &=&
-2\, \frac{\Gamma(1-\veps)\Gamma(-\veps-\al) \Gamma(\al+2\veps)}{\Gamma(2-2\veps)} 
\label{KotikovG}\\
&\times& \left[ \frac{\Gamma(1-\veps)}{\Gamma(1-\al-3\veps)}\,
\sum_{n=0}^{\infty}\,\frac{\Gamma(n+2-2\veps)}{\Gamma(n+2+\al)}\,\frac{1}{n+1+\al+\veps}
- \pi \cot[\pi(\al+2\veps)] \right]\, .
\nonum
\eea
As it was discussed already in Ref.~[\onlinecite{Kotikov:1995cw}], the equality of the two results (\ref{Kazakov}) and (\ref{KotikovG}) provides the following
relation between two ${}_3F_2$-hypergeometric functions of argument $-1$ and a single  ${}_3F_2$-hypergeometric function of argument $1$:
%
\begin{flalign}
&{}_3F_2(2A,B,1;B+1,2-A;-1)+ \frac{B}{1+A-B} \, {}_3F_2(2A,1+A-B,1;2+A-B,2-A;-1) \nonumber \\
&= B \cdot \frac{\Gamma(2-A)\Gamma(B+A-1)\Gamma(B-A)\Gamma(1+A-B)}{\Gamma(2A)\Gamma(1+B-2A)} - \frac{1-A}{B+A-1} \, {}_3F_2(2A,B,1;B+1,A+B;1) \, ,
\label{F32}
\end{flalign}
%
where, hereafter $A$, $B$ and $C$ are arbitrary. Notice that such a relation does not appear in standard textbooks.

The purpose of the present short paper is to demonstrate {\it analytically} the equality of the two above results (\ref{Kazakov}) and (\ref{KotikovG})
and, hence, to recover analytically the relation (\ref{F32}) between ${}_3F_2$-hypergeometric functions.
The demonstration will be given in details as the method of calculation involved, or some generalization of it, may be useful to other physicists and mathematicians.
Actually, the use of integral representations as done below is rather similar to manipulations of $\Psi$-functions and 
associated functions carried out earlier [\onlinecite{Kotikov:2000pm,Kotikov:2002ab}] in order to simplify the famous
Balitsky-Fadin-Kuraev-Lipatov (BFKL) equation at the first two orders of perturbation theory.

As a by product of our analysis, we obtain several other representations for the considered diagram $I(\al)$ which, in-turn, 
should be useful for its high-order $\veps$-expansion in the case where $\alpha=1+a\veps$.
Such expansions are planned for our future investigations.

\subsection{Particular cases}

Before closing this section, it is convenient to consider two particular cases: $\al=0$ and $\al=-\veps$, 
where $I(1+\al)$ can be represented in terms of a combination of $\Gamma$-functions.

In the case  $\al=0$, an application of the IBP procedure yields:
\bea
I(1) &=& \frac{1}{\veps^3} \, \frac{\Gamma^4(1-\veps)\Gamma^2(1+\veps)}{(1-2\veps)\Gamma^2(1-2\veps)}
\left[1 -  \frac{\Gamma^2(1-2\veps)\Gamma(1+2\veps)}{\Gamma^2(1+\veps)\Gamma(1-\veps)\Gamma(1-3\veps)} \right] \nonumber \\
 &=& \frac{1}{\veps} \, \frac{\Gamma^2(1-\veps)}{(1-2\veps)\Gamma^2(1-2\veps)} \, \frac{\pi^2}{\sin^2[\pi\veps]} \,
\left[1 -  \frac{\Gamma(1-2\veps)}{\Gamma(1+\veps)\Gamma(1-3\veps)} \, \frac{1}{\cos[\pi\veps]}
\right] \, .
\label{al=0}
\eea

In the case $\al=-\veps$, the diagram has been calculated by the method of uniqueness (see Ref.~[\onlinecite{Vasiliev:1981dg}] and also discussions in Refs.~[\onlinecite{KivelSV93,VassilievBook}]), 
and reads:
\be
I(1-\veps) = 3\,\frac{\Gamma(1-\veps)\Gamma(\veps)}{\Gamma(2-2\veps)} \,\Big[ \psi'(1-\veps) - \psi'(1) \Big] \, .
\label{result:I}
\ee
A modern evaluation of the result can be found in Ref.~[\onlinecite{Kotikov:2013kcl}].
Notice that, with respect to practical applications, the result (\ref{result:I}) has been used in Ref.~[\onlinecite{Kotikov:2013kcl}]
in order to recover some of the results of Ref.~[\onlinecite{Teber:2012de}] where the more general Eq.~(\ref{KotikovG}) was used
in order to compute the most complicated part of the two-loop correction to electromagnetic current correlations in brane worlds / planar condensed
matter physics systems.

\section{Transformation of Eq.~(\ref{Kazakov})}  
\label{sec:2}

In this section, we shall demonstrate analytically how the result of Eq.~(\ref{KotikovG}) can be obtained from the result of Eq.~(\ref{Kazakov}).

As a starting point, let us note that Kazakov's formula (\ref{Kazakov}) was initially written under the following form:
\bea
I(1+\al) &=&
-2\, \frac{\Gamma^2(1-\veps)\Gamma(\veps)\Gamma(-\veps-\al) \Gamma(\al+2\veps)}{\Gamma(2-2\veps)\Gamma(1+\al)\Gamma(1-3\veps-\al)} 
\label{KazakovIni}\\
&\times& 
\Biggl[
\sum_{n=1}^{\infty}\,(-1)^n \frac{\Gamma(n+1-2\veps)}{\Gamma(n+\veps)}\,\left(\frac{1}{n+\al+\veps} + \frac{1}{n-\al-2\veps}\right)
\nonum \\
&&-\cos [\pi \veps] \times 
\sum_{n=1}^{\infty}\,(-1)^n \frac{\Gamma(n+1-3\veps)}{\Gamma(n)}\,\left(\frac{1}{n+\al} + \frac{1}{n-\al-3\veps}\right)
\Biggr]\, ,
\nonum
\eea
and that the last sum in the rhs is equal to: $-\Gamma(1+\al)\Gamma(1-3\veps-\al)$.

\subsection{Transformation of the last term in Eq.~(\ref{KazakovIni})} 

It is instructive to first evaluate the last term in Eq.~(\ref{KazakovIni}). In order to do so, we can represent the 
factor $(1/(n+\al)+1/(n-\al-3\veps))$ in the following integral form:
\be
\frac{1}{n+\al} + \frac{1}{n-\al-3\veps} = \int^1_0 dx \, x^{n-1} \left[ x^{\al} + x^{-\al-3\veps} \right] \, .
\ee
Then, the last sum in the rhs of Eq.~(\ref{KazakovIni}) can be summed as:
\be
\sum_{n=1}^{\infty}\,(-1)^n \frac{\Gamma(n+1-3\veps)}{\Gamma(n)}\, x^{n-1} = - \frac{\Gamma(2-3\veps)}{(1+x)^{2-3\veps}} \, ,
\ee
which yields:
\be
\sum_{n=1}^{\infty}\,(-1)^n \frac{\Gamma(n+1-3\veps}{\Gamma(n)}\,\left(\frac{1}{n+\al} + \frac{1}{n-\al-3\veps}\right)
= -\Gamma(2-3\veps) \int^1_0 dx \, \frac{\left[ x^{\al} + x^{-\al-3\veps} \right]}{(1+x)^{2-3\veps}} \, .
\label{int}
\ee
After the replacement $x \to y=1/x$, the last term in Eq.~(\ref{int}) reads:
\be
\int^1_0 dx \, \frac{x^{-\al-3\veps}}{(1+x)^{2-3\veps}} = \int^{\infty}_1 dy \, \frac{y^{\al}}{(1+y)^{2-3\veps}}\, .
\ee
The result (\ref{int}) can then be expressed as:
\be
\sum_{n=1}^{\infty}\,(-1)^n \frac{\Gamma(n+1-3\veps)}{\Gamma(n)}\,\left(\frac{1}{n+\al} + \frac{1}{n-\al-3\veps}\right)
= -\Gamma(2-3\veps) \int^{\infty}_0 dx \, \frac{x^{\al}}{(1+x)^{2-3\veps}} \, ,
\label{int.1}
\ee
where the last integral is evaluated in terms of $\Gamma$-functions because:
\be
\int^{\infty}_0 dx \, \frac{x^{\al}}{(1+x)^{\beta}} = \frac{\Gamma(\al+1) \Gamma(\beta-\al-1)}{\Gamma(\beta)}\, .
\ee
We therefore come to the final result having the following advertised form:
\be
\sum_{n=1}^{\infty}\,(-1)^n \frac{\Gamma(n+1-3\veps)}{\Gamma(n)}\,\left(\frac{1}{n+\al} + \frac{1}{n-\al-3\veps}\right)
= -\Gamma(1+\al)\Gamma(1-3\veps-\al)\, .
\label{int.2}
\ee

\subsection{Transformation of the first sum in the rhs of Eq.~(\ref{KazakovIni}) to an ${}_3F_2$ function of argument $1$} 

Following the analysis of the previous subsection, it is convenient 
to represent the factor $(1/(n+\al+\veps)+1/(n-\al-2\veps))$ in the following form:
\be
\frac{1}{n+\al+\veps} + \frac{1}{n-\al-2\veps} = \int^1_0 dx \, x^{n-1} \left[ x^{\al+\veps} + x^{-\al-2\veps} \right] \, .
\ee
The first series in the rhs of Eq.~(\ref{KazakovIni}) can then be summed as:
\bea
\sum_{n=1}^{\infty}\,(-1)^n \frac{\Gamma(n+1-3\veps)}{\Gamma(n+\veps)}\, x^{n-1} &=& -\frac{\Gamma(2-2\veps)}{\Gamma(1+\veps)}
\, {}_2F_1(2-2\veps,1;1+\veps;-x) \nonumber \\
&=&  -\frac{\Gamma(2-2\veps)}{\Gamma(1+\veps)} \, \frac{1}{(1+x)^{2-3\veps}} \, 
{}_2F_1(3\veps-1,\veps;1+\veps;-x)\, ,
\nonumber 
 \eea
where we have used the standard transformation formula for the ${}_2F_1$-hypergeometric function:
\be
{}_2F_1(A,B;C;z) = (1-z)^{C-A-B} \, {}_2F_1(C-A,C-B;C;z)\, .
\ee
So, the first sum in the rhs of Eq.~(\ref{KazakovIni})  can be represented as:
\bea
&&J_1 \equiv - 
\sum_{n=1}^{\infty}\,(-1)^n \frac{\Gamma(n+1-2\veps)}{\Gamma(n+\veps)}\,\left(\frac{1}{n+\al+\veps} + \frac{1}{n-\al-2\veps}\right)
\nonumber \\
&&= 
\frac{\Gamma(2-2\veps)}{\Gamma(1+\veps)} \int^1_0 dx \,\frac{ x^{\al+\veps} + x^{-\al-2\veps}}{(1+x)^{2-3\veps}} \,
{}_2F_1(3\veps-1,\veps;1+\veps;-x)\, \nonumber \\
&& = \frac{\Gamma(2-2\veps)}{\Gamma(\veps)} \int^1_0 dx \,\frac{ x^{\al+\veps} + x^{-\al-2\veps}}{(1+x)^{2-3\veps}} \,
\int^1_0 dt \, \frac{t^{\veps-1}}{(1+xt)^{3\veps-1}} \equiv \frac{\Gamma(2-2\veps)}{\Gamma(\veps)} \, \tilde{J}_1  \, ,
\label{Int}
\eea
where we have used the integral representation of the ${}_2F_1$-hypergeometric function:
\be
{}_2F_1(A,B;C;z) = \frac{\Gamma(C)}{\Gamma(B)\Gamma(C-B)} \, \int^1_0 dt \, 
\frac{t^{B-1} (1-t)^{C-B-1}}{(1-zt)^A} \, .
\label{Integral}
\ee

After some algebra (see Appendix A) we have:
\bea
J_1 &=& \frac{\Gamma(2-2\veps)}{\Gamma(1+\veps)} \,  \int^{1}_{0} dx_2 \,\frac{(1-x_2)^{\al}}{x_2^{\al+2\veps}} \, 
{}_2F_1\left(-\al,\veps;1+\veps;\frac{x_2}{x_2-1}\right)  
 \nonumber \\ 
&=& \frac{\Gamma(2-2\veps)}{\Gamma(1+\veps)} \,  \int^{1}_{0} dx_2 \,\frac{1}{x_2^{\al+2\veps}} \, 
{}_2F_1(-\al,1;1+\veps;x_2) \, ,
\label{Int.3}
\eea
where we have used another standard transformation formula for the ${}_2F_1$-hypergeometric function:
\be
{}_2F_1(A,B;C;z) = (1-z)^{-A} \, {}_2F_1\left(A,C-B;C;\frac{z}{z-1}\right)\, .
\ee
Taking the standard series representation for the ${}_2F_1$-hypergeometric functions, we have for  $J_1$ the following final result:
\be
J_1 =   \int^{1}_{0} dx_2 \,\frac{1}{x_2^{\al+2\veps}} \, \sum_{n=0}^{\infty} \frac{\Gamma(n-\al)\Gamma(2-2\veps)}{\Gamma(-\al)\Gamma(n+1+\veps)}
x_2^{n} = \sum_{n=0}^{\infty} \frac{\Gamma(n-\al)\Gamma(2-2\veps)}{\Gamma(-\al)\Gamma(n+1+\veps)} \, \frac{1}{n+1-2\veps-\al} \, ,
\label{int.3}
\ee
which is expressed in terms of a single  ${}_3F_2$-hypergeometric function of argument $1$. 

However, the result does not coincide yet with the one of Eq.~(\ref{KotikovG}). 
Indeed, we have now:
\bea
I(1+\al) &=&
2\, \frac{\Gamma^2(1-\veps)\Gamma(\veps)\Gamma(-\veps-\al) \Gamma(\al+2\veps)}{\Gamma(2-2\veps)} \Biggl[ 
\frac{1}{\Gamma(1+\al)\Gamma(1-3\veps-\al)}
\label{Kazakov1}\\
&\times& 
 \sum_{n=0}^{\infty} \frac{\Gamma(n-\al)\Gamma(2-2\veps)}{\Gamma(-\al)\Gamma(n+1+\veps)} \, \frac{1}{n+1-2\veps-\al} 
-\cos [\pi \veps] \Biggr]\, ,
\nonum
\eea
where the series as well as the term containing only $\Gamma$-functions
differ from  the corresponding ones in Eq.~(\ref{KotikovG}).

\subsection{Transformation of the series in the rhs of Eq.~(\ref{Kazakov1}) to the one in Eq.~(\ref{KotikovG})} 

We now use the transformation formula for the  ${}_3F_2$-hypergeometric function of argument $1$:~\cite{Bailey}
\bea
&&{}_3F_2(A,B,C;E,F;1) = \frac{\Gamma(1-A)\Gamma(E)\Gamma(F)\Gamma(C-B)}{\Gamma(E-B)\Gamma(F-B)\Gamma(1+B-A)\Gamma(C)} \nonumber \\
&& \times {}_3F_2(B,B-E+1,B-F+1;B-C+1,B-A+1;1) + \Bigl( B \leftrightarrow  C \Bigr)\, .
\label{Trans}
\eea
For $E=B+1$, the ${}_3F_2$-hypergeometric function can be represented as the sum of another ${}_3F_2$-hypergeometric function and a term
containing only $\Gamma$-functions. In terms of series representations, the relation has the following form:
\bea
&& \sum_{n=0}^{\infty} \frac{\Gamma(n+A)\Gamma(n+C)}{n!\Gamma(n+F)} \, \frac{1}{n+B} =  
\frac{\Gamma(1-A)\Gamma(A)\Gamma(B)\Gamma(C-B)}{\Gamma(F-B)\Gamma(1+B-A)} \nonumber \\
&& - \frac{\Gamma(1-A)\Gamma(A)}{\Gamma(F-C)\Gamma(1+C-F)} \, \sum_{n=0}^{\infty} \frac{\Gamma(n+C-F+1)\Gamma(n+C)}{n!\Gamma(n+C-A+1)} \, 
\frac{1}{n+C-B} \, .
\label{Tran1}
\eea
Using this relation with:
$B=1-\al - 2\veps$, $F=1+\veps$,  $A= - \al$ and  $C=1$,
we can transform the series in Eq.~(\ref{Kazakov1}) as follows:
\bea
&& \sum_{n=0}^{\infty} \frac{\Gamma(n-\al)\Gamma(2-2\veps)}{\Gamma(-\al)\Gamma(n+1+\veps)} \, \frac{1}{n+1-2\veps-\al}
= \frac{\Gamma(1+\al)\Gamma(1-\al-2\veps)\Gamma(\al+2\veps)}{\Gamma(\al+3\veps)} \nonumber \\
&& - \frac{\Gamma(1+\al)\Gamma(2-2\veps)}{\Gamma(\veps)\Gamma(1-\veps)}  
\sum_{n=0}^{\infty} \frac{\Gamma(n+1-\veps)}{\Gamma(n+2+\al)} \, \frac{1}{n+2\veps+\al}\, .
\nonumber 
\eea
Focusing on the terms containing only products of $\Gamma$-functions, they can be simplified as follows:
\bea
&&\frac{\Gamma(1-\al-2\veps)\Gamma(\al+2\veps)}{\Gamma(1-\al-3\veps)\Gamma(\al+3\veps)} - \cos[\pi \veps] =
\frac{\sin[\pi(\al+3\veps)]}{\sin[\pi(\al+2\veps)]} - \cos[\pi \veps]  \nonumber \\
&&= \frac{\sin[\pi\veps]\cos[\pi(\al+2\veps)]}{\sin[\pi(\al+2\veps)]}
 =  \sin[\pi\veps] \cot[\pi(\al+2\veps)] \, .
\nonumber 
\eea
Then, $I(1+\al)$ can be written as:
\bea
I(1+\al) &=&
2\, \frac{\Gamma(1-\veps)\Gamma(-\veps-\al) \Gamma(\al+2\veps)}{\Gamma(1-\al-3\veps)} \Biggl[ 
-\sum_{n=0}^{\infty} \frac{\Gamma(n+1-\veps)}{\Gamma(n+2+\al)} \, \frac{1}{n+2\veps+\al}
\label{Kazakov2}\\
&+& \frac{\Gamma(1-\al-3\veps)}{\Gamma(2-2\veps)} \, \pi \cot [\pi(\al+2 \veps)]
\Biggr]\, .
\nonum
\eea
At this point, we can see that the terms containing only products of $\Gamma$-functions are the same in Eqs.~(\ref{KotikovG}) and (\ref{Kazakov2}). 
Hence,  the corresponding series should be identical too. To demonstrate this, it is convenient to use the following  transformation formula for the 
${}_3F_2$-hypergeometric function of argument $1$:~\cite{Bailey}
\begin{flalign}
&&{}_3F_2(A,B,C;E,F;1) = \frac{\Gamma(E)\Gamma(F)\Gamma(S)}{\Gamma(A)\Gamma(S+B)\Gamma(S+C)} 
{}_3F_2(E-A,F-A,S;S+B,S+C;1)\, ,
\label{Trans.1}
\end{flalign}
where $S=E+F-A-B-C$. Noticing that the sum in the rhs of Eq.~(\ref{Kazakov2}) has the form:
\be
\sum_{n=0}^{\infty}\,\frac{\Gamma(n+1-\veps)}{\Gamma(n+2+\al)}\,\frac{1}{n+\al+2\veps} = \frac{\Gamma(1-\veps)}{\Gamma(2+\al)} \, \frac{1}{\al+2\veps} \,
{}_3F_2(1-\veps, 1,\al+2\veps;2+\al,1+\al+2\veps;1) \, ,
\ee
and applying Eq.~(\ref{Trans.1}) with $A=\al+2\veps$, yields:
\be
{}_3F_2(1-\veps, 1,\al+2\veps;2+\al,1+\al+2\veps;1) = \frac{(\veps+2\al)}{(1+\al+\veps)} {}_3F_2(2-2\veps, 1,1+\al+\veps;2+\al,2+\al+\veps;1)\, .
\ee
Hence:
\be
\sum_{n=0}^{\infty} \frac{\Gamma(n+1-\veps)}{n!\Gamma(n+2+\al)} \, \frac{1}{n+2\veps+\al}
 =
 \frac{\Gamma(1-\veps)}{\Gamma(2-2\veps)} \, 
 \sum_{n=0}^{\infty}\,\frac{\Gamma(n+2-2\veps)}{\Gamma(n+2+\al)}\,\frac{1}{n+1+\al+\veps}\, ,
\ee
which proves the equality of the rhs of Eqs.~(\ref{KotikovG}) and (\ref{Kazakov2}).

\section{Other useful representations for $I(1+\al)$}

Eqs.~(\ref{Kazakov1}) and (\ref{Kazakov2}) can be considered as new results for the diagram $I(1+\al)$.
Other new results can be obtained with the help of the transformation formula (\ref{Trans}) with $A=1-\delta$ and $\delta \to 0$; the later will involve the replacement
of a $\Gamma$-function by a $\Psi$-function which may be more convenient to expand the diagram in $\veps$. 
Notice that similar representations, but for a two-loop massless Feynman diagram with two arbitrary non-adjacent indices,
have been obtained in our paper [\onlinecite{Kotikov:2013eha}], see App.~\ref{app2} for a summary of our results.

In order to proceed with the derivation, we consider the sum:
\be
\sum_{n=0}^{\infty}\,\frac{\Gamma(n+2-2\veps)}{\Gamma(n+2+\al)}\,\frac{\Gamma(n+1-\delta)}{n!\Gamma(1-\delta)}\,
\frac{1}{n+1+\al+\veps} \, ,
\nonumber
\ee
which coincides with the one  in Eq.~(\ref{KotikovG}) when $\delta \to 0$. Applying the transformation formula Eq.~(\ref{Trans})  with
$A=1-\delta$, $C = 2-2\veps$, $F= 2+\al$  and  $B= 1+\al+\veps$,
yields:
\bea
&&\sum_{n=0}^{\infty}\,\frac{\Gamma(n+2-2\veps)}{\Gamma(n+2+\al)}\,\frac{\Gamma(n+1-\delta)}{n!\Gamma(1-\delta)}\,
\frac{1}{n+1+\al+\veps} = \frac{\Gamma(\delta)\Gamma(1-\al-3\veps)\Gamma(1+\al+\veps)}{\Gamma(1-\veps)\Gamma(1+\al+\veps+\delta)}\, \nonumber \\
&& - \frac{\Gamma(\delta)}{\Gamma(1-\al-2\veps)\Gamma(\al+2\veps)}\, 
\sum_{n=0}^{\infty}\,\frac{\Gamma(n+1-\al-2\veps)\Gamma(n+2-2\veps)}{n!\Gamma(n+2-2\veps+\delta)} \, \frac{1}{n+1-\al-3\veps}\, .
\nonumber 
\eea
Taking the limit  $\delta \to 0$, we can transform the rhs to the following form:
\bea
&& \frac{\Gamma(1-\al-3\veps)}{\Gamma(1-\veps)} \, \left[ \frac{1}{\delta} + \Psi(1) - \Psi(1+\al+\veps) \right]
-  \frac{1}{\Gamma(1-\al-2\veps)\Gamma(\al+2\veps)}\, 
  \nonumber \\
&&\times \sum_{n=0}^{\infty}\,\frac{\Gamma(n+1-\al-2\veps)}{n!} \, \frac{1}{n+1-\al-3\veps}
\, \left[ \frac{1}{\delta} + \Psi(1) - \Psi(n+2-2\veps) \right] + O(\delta)\, .
\nonumber 
\eea
The series appearing in factor of $1/\delta$ can be summed as:
\be
 \sum_{n=0}^{\infty}\,\frac{\Gamma(n+1-\al-2\veps)}{n!} \, \frac{1}{n+1-\al-3\veps} = 
\frac{\Gamma(1-\al-2\veps)\Gamma(\al+2\veps)\Gamma(1-\al-3\veps)}{\Gamma(1-\veps)} \, ,
\nonumber 
\ee
which leads to the cancellation of all terms $\sim 1/\delta$. Hence, we obtain:
\bea
&&\sum_{n=0}^{\infty}\,\frac{\Gamma(n+2-2\veps)}{\Gamma(n+2+\al)}\, \frac{1}{n+1+\al+\veps} = 
 - \frac{\Gamma(1-\al-3\veps)}{\Gamma(1-\veps)} \,  \Psi(1+\al+\veps)
  \nonumber \\
&&+  \frac{1}{\Gamma(1-\al-2\veps)\Gamma(\al+2\veps)}\, \sum_{n=0}^{\infty}\,\frac{\Gamma(n+1-\al-2\veps)}{n!} \, 
\frac{\Psi(n+2-2\veps)}{n+1+\al-3\veps} \, .
\nonumber
\eea

Similar calculations can be repeated for the series appearing in the rhs of Eqs.~(\ref{Kazakov1}) and (\ref{Kazakov2}). The corresponding results read:
\bea 
&&
\sum_{n=0}^{\infty} \frac{\Gamma(n-\al)}{\Gamma(n+1+\veps)} \, \frac{1}{n+1-2\veps-\al} =
-  \frac{\Gamma(2\veps-1)}{\Gamma(\al+3\veps)} \, \Psi(1-\al-2\veps) 
 \nonumber \\
&&+  \frac{1}{\Gamma(-\al-\veps)\Gamma(1+\al+\veps)}\, \sum_{n=0}^{\infty}\,\frac{\Gamma(n-\al-\veps)}{n!} \, \frac{\Psi(n-\al)}{n+2\veps-1}\, ,
\nonumber
\eea
and
\bea
&&\sum_{n=0}^{\infty}\,\frac{\Gamma(n+1-\veps)}{\Gamma(n+2+\al)}\, \frac{1}{n+\al+2\veps} = 
-  \frac{\Gamma(1-\al-3\veps)}{\Gamma(2-2\veps)} \, \Psi(\al+2\veps) 
  \nonumber \\
&&+  \frac{1}{\Gamma(1+\al+\veps)\Gamma(-\al-\veps)}\, \sum_{n=0}^{\infty}\,\frac{\Gamma(n-\al-\veps)}{n!} \, 
\frac{\Psi(n+1-\veps)}{n+1-\al-3\veps} \, .
\nonumber
\eea

\section{Conclusion}

We have analytically proven the equality of two previously existing results (\ref{Kazakov}) and (\ref{KotikovG}), containing 
 ${}_3F_2$-hypergeometric functions of arguments $-1$ and $1$, respectively. As a by-product of our calculations, the relation 
(\ref{F32}) between these two types of  ${}_3F_2$-hypergeometric functions has been proven exactly.
Such a relation should be useful for various applications because there are a lot of transformation formulas available for the ${}_3F_2$-hypergeometric function of argument
$1$ (see [\onlinecite{Bailey}]). Thus, Eq.~(\ref{F32}) gives a possibility to apply some of these  transformation formulas to  
the ${}_3F_2$-hypergeometric function of argument $-1$. Moreover, as already mentioned in the introduction, the way
integral representations were used in Sec.~\ref{sec:2} and App.~\ref{app} is rather similar to manipulations
carried out with $\Psi$-functions and associated functions in [\onlinecite{Kotikov:2000pm,Kotikov:2002ab}] in order to simplify
BFKL results at the first two orders of perturbation theory and to verify their conformal properties or a violation of them.

  
Moreover, using transformation formulas  Eqs.~(\ref{Trans}) and (\ref{Trans.1}) for ${}_3F_2$-hypergeometric functions
of argument $1$ [\onlinecite{Bailey}]
we have found several new representations for the diagram $I(\al)$ under consideration.
 We hope to use them in our future $\veps$-expansions of $I(\al)$ and $J(\al_1,\al_2,\al_3,\al_4,\al_5)$ with values
of the indices $\al$ and $\al_i$ $(i=1,..., 5)$ close to $1$.

\begin{figure}[!tbp]
  \centering
  \subfloat[$J(1/2,1,1/2,\al,1)$]{\includegraphics{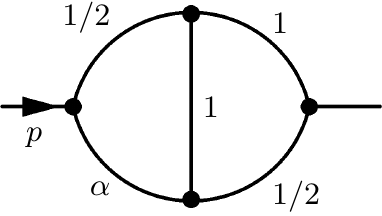}\label{fig:3}}
  \qquad \qquad \qquad \qquad
  \subfloat[$J(1/2,1,1/2,1,\al)$]{\includegraphics{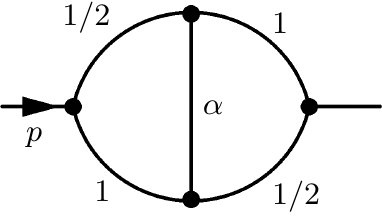}\label{fig:4}}
  \caption{Examples of complicated diagrams appearing in Refs.~[\onlinecite{Kotikov:2016wrb,Kotikov:2016prf,Kotikov:2016yrn}].}
\end{figure}

In closing, our analysis was devoted to the simplest non-trivial two-loop massless propagator-type Feynman diagram. 
We hope that such an analysis may be generalized to allow the exact computation of more complicated 
diagrams such as those appearing upon studying $1/N$ corrections 
to dynamical fermion mass generation in QED$_3$~\cite{Kotikov:2016wrb,Kotikov:2016prf} and RQED$_{4,3}$~\cite{Kotikov:2016yrn}, see Figs.~\ref{fig:3}
and \ref{fig:4}.
 
\acknowledgments

One of us (A.V.K.) was supported by RFBR grant 16-02-00790-a. 
Financial support from Universit\'e Pierre et Marie Curie and CNRS is acknowledged.

\appendix

\section{Evaluation of the integral $\tilde{J}_1$}
\label{app}

In this Appendix, we give the details of the evaluation of the integral $\tilde{J}_1$ in Eq.~(\ref{Int}).

Introducing the new variable $t_1=xt$, we have for $\tilde{J}_1$:
\be
\tilde{J}_1 = \int^1_0 dx \,\frac{ x^{\al} + x^{-\al-3\veps}}{(1+x)^{2-3\veps}} \, \int^x_0 dt_1 \, \frac{t_1^{\veps-1}}{(1+t_1)^{3\veps-1}}
=  \int^1_0 dt_1 \, \frac{t_1^{\veps-1}}{(1+t_1)^{3\veps-1}} \, \int^1_{t_1} dx \,\frac{ x^{\al} + x^{-\al-3\veps}}{(1+x)^{2-3\veps}}\, .
\label{app:Int.1}
\ee
Note that after the replacement $x \to y=1/x$ the last term in the last integral of (\ref{app:Int.1}) reads:
\be
\int^1_{t_1} dx \, \frac{x^{-\al-3\veps}}{(1+x)^{2-3\veps}} = \int^{1/t_1}_1 dy \, \frac{y^{\al}}{(1+y)^{2-3\veps}}\, ,
\ee
and, thus, the result (\ref{app:Int.1}) can be expressed as:
\be
\tilde{J}_1 = \int^1_0 dt_1 \, \frac{t_1^{\veps-1}}{(1+t_1)^{3\veps-1}} \, \int^{1/t_1}_{t_1} dx \,\frac{x^{\al}}{(1+x)^{2-3\veps}}\, .
\label{app:int.2}
\ee
Using the new variables $x_1=1/(1+x)$, and later $x_1=x_2/(1+t_1)$, we have for the last integral in (\ref{app:int.2}):
\be
\int^{1/t_1}_{t_1} dx \,\frac{x^{\al}}{(1+x)^{2-3\veps}} = \int^{1/(1+t_1)}_{t_1/(1+t_1)} dx_1 \,\frac{(1-x_1)^{\al}}{x_1^{\al+3\veps}}
 = \frac{1}{(1+t_1)^{1-\al-3\veps}} \, \int^{1}_{t_1} dx_2 \,\frac{\left(1-\frac{x_2}{1+t_1}\right)^{\al}}{x_2^{\al+3\veps}}\, .
\label{app:int.3}
\ee
Substituting the new result (\ref{app:int.3}) to the rhs of (\ref{app:int.2}), we obtain:
%
\be
\tilde{J}_1 = \int^1_0 dt_1 \, \frac{1}{t_1^{1-\veps}} \,  \int^{1}_{t_1} dx_2 \,\frac{(1+t_1-x_2)^{\al}}{x_2^{\al+3\veps}} =
\int^{1}_{0} dx_2 \,\frac{1}{x_2^{\al+3\veps}} \, \int^{x_2}_0 dt_1 \, \frac{(1+t_1-x_2)^{\al}}{t_1^{1-\veps}}\, .
\ee
Using the new variable $t_2=t_1/x_2$,  we see that:
\bea
&&\tilde{J}_1 =  \int^{1}_{0} dx_2 \,\frac{1}{x_2^{\al+2\veps}} \, \int^{1}_0 dt_2 \, \frac{(1-x_2+x_2t_2)^{\al}}{t_1^{1-\veps}}
= \int^{1}_{0} dx_2 \,\frac{(1-x_2)^{\al}}{x_2^{\al+2\veps}} \, \int^{1}_0 dt_2 \, \frac{\left(1+\frac{x_2t_2}{1-x_2}\right)^{\al}}{t_2^{1-\veps}}
 \nonumber \\ &&
= \frac{\Gamma(\veps)}{\Gamma(1+\veps)} \,  \int^{1}_{0} dx_2 \,\frac{(1-x_2)^{\al}}{x_2^{\al+2\veps}} \, 
{}_2F_1\left(-\al,\veps;1+\veps;\frac{x_2}{x_2-1}\right)\, , 
\eea
where, in the last step, we have applied the property (\ref{Integral}).

\section{Reminder of results for $J(\al,1,\beta,1,1)$}
\label{app2}

In this appendix, we recall, for completeness, the exact results obtained in 
Ref.~[\onlinecite{Kotikov:2013eha}], based on the results of Ref.~[\onlinecite{Kotikov:1995cw}],  for the diagram $J(\al,1,\beta,1,1)$ with two arbitrary non-adjacent indices, see Fig.~\ref{fig:5}.
As will be seen below, these results show that $J(\al,1,\beta,1,1)$ can be expressed in terms of a linear combination of two ${}_3F_2$-hypergeometric functions of argument $1$. 
Moreover, as in the case of $J(\al)$ above, some representations involve $\Psi$-functions.

\begin{figure}[!tbp]
  \centering
   \includegraphics{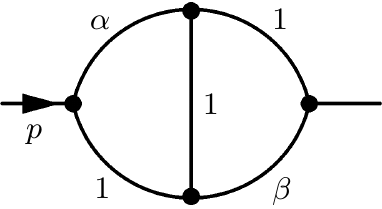}
   \caption{\label{fig:5}
   The diagram $J(\al,1,\beta,1,1)$ with two arbitrary non-adjacent indices.}
\end{figure}

To start with, we extract the momentum dependence of the diagram which leads to:
\be
J(\al,1,\beta,1,1) = \frac{p^{2\,d_F}}{(4\pi)^D}\,G(\al,1,\beta,1,1)\, ,
\ee
where $d_F = D - 3 -\al -\beta = 1 -\al -\beta -2\veps$ and we use the notation $G(\al,1,\beta,1,1)$ for the dimensionless coefficient function.
The later reads:~\cite{Kotikov:2013eha}
\be
G(\al,1,\beta,1,1) =
\frac{1}{\tilde{\al}-1} \frac{1}{1-\tilde{\beta}} \,
\frac{\Gamma(\tilde{\al})\Gamma(\tilde{\beta})
\Gamma(3-\tilde{\al}-\tilde{\beta})}{\Gamma(\al)
\Gamma(\lambda-2+\tilde{\al}+\tilde{\beta})}
\frac{\Gamma(\lambda)}{\Gamma(2\lambda)} \, I(\tilde{\al},\tilde{\beta}) \, ,
\label{Gab}
\ee
where $\tilde{\al} = D/2-\al$, $\lambda=D/2-1$, $D=4-2\veps$. In Eq.~(\ref{Gab}), the function $I(\tilde{\al},\tilde{\beta})$ can be written in four different forms
which read:~\cite{Kotikov:2013eha}
\bea
&&I(\tilde{\al},\tilde{\beta}) =
\frac{\Gamma(1+\lambda-\tilde{\al})}{
\Gamma(3-\tilde{\al}-\tilde{\beta})}
\frac{\pi \sin[\pi(\tilde{\beta}- \tilde{\al}+\lambda)]}{\sin[\pi (\lambda-1+\tilde{\beta})]\sin[\pi \tilde{\al}]}
+ \sum_{n=0}^{\infty} \frac{\Gamma(n+2\lambda)}{n!(n+\lambda+\tilde{\al}-1)}
\label{I1} \\
&& \times \biggl( \frac{\Gamma(n+1)}{\Gamma(n+2+\lambda-\tilde{\beta})}
- \frac{\Gamma(n-2+\lambda +\tilde{\al}+\tilde{\beta})\Gamma(2-\tilde{\beta}) \Gamma(\lambda)
}{\Gamma(n-1+2\lambda+\tilde{\al})\Gamma(3-\tilde{\al}-\tilde{\beta})\Gamma(\lambda+\tilde{\al}-1)}
\frac{\sin[\pi(\tilde{\beta}+\lambda-1)]}{\sin[\pi \tilde{\al}]}
\biggl)
\, ,
\nonum
\eea
\bea
&&I(\tilde{\al},\tilde{\beta}) =
\frac{\Gamma(1+\lambda - \tilde{\al})}{
\Gamma(3-\tilde{\al}-\tilde{\beta})}
\frac{\pi \sin[\pi \tilde{\al}]}{\sin[\pi (\lambda-1+\tilde{\beta})] \sin[\pi (\tilde{\al}+\tilde{\beta}+\lambda-1) ]}
\label{I2} \\&&
+ \sum_{n=0}^{\infty} \frac{\Gamma(n+2\lambda)}{n!
}
\biggl(\frac{1}{n+\lambda+\tilde{\al}-1}
\frac{\Gamma(n+1)}{\Gamma(n+2+\lambda-\tilde{\beta})}
\nonumber \\&&
+ \frac{1}{n+\lambda+1-\tilde{\al}}
\frac{\Gamma(n+2-\tilde{\al})\Gamma(2-\tilde{\beta})\Gamma(\lambda)
}{\Gamma(n+3+\lambda-\tilde{\al}-\tilde{\beta})\Gamma(3-\tilde{\al}-\tilde{\beta})\Gamma(\lambda+\tilde{\al}-1)}
\frac{\sin[\pi(\tilde{\beta}+\lambda-1)]}{\sin[\pi (\tilde{\al}+\tilde{\beta}+\lambda-1)]}
\biggl)
\, ,
\nonum
\eea
\bea
&&I(\tilde{\al},\tilde{\beta}) =
\frac{\Gamma(1+\lambda-\tilde{\al})}{
\Gamma(3-\tilde{\al}-\tilde{\beta})}
\frac{\pi \sin[\pi(\tilde{\beta}- \tilde{\al}+\lambda)]}{\sin[\pi (\lambda-1+\tilde{\beta})]\sin[\pi \tilde{\al}]}
- \frac{\Gamma(1+\lambda-\tilde{\al})}{\Gamma(3-\tilde{\al}-\tilde{\beta})}
\Psi(\lambda+\tilde{\al}-1)
\label{I3} \\
&&+ \sum_{n=0}^{\infty} \frac{\Gamma(n+\tilde{\beta}+\lambda-1)}{n!
(n+\lambda+1-\tilde{\al})} \Psi(n+2\lambda) \frac{\sin[\pi(\tilde{\beta}+\lambda-1)]}{\pi }
\nonumber \\
&&
- \sum_{n=0}^{\infty} \frac{\Gamma(n+2\lambda)}{n!(n+\lambda+\tilde{\al}-1)}
\frac{\Gamma(n-2+\lambda +\tilde{\al}+\tilde{\beta})\Gamma(2-\tilde{\beta})\Gamma(\lambda)
}{\Gamma(n-1+2\lambda+\tilde{\al})\Gamma(3-\tilde{\al}-\tilde{\beta})\Gamma(\lambda+\tilde{\al}-1)}
\frac{\sin[\pi(\tilde{\beta}+\lambda-1)]}{\sin[\pi \tilde{\al}]}
\nonum
\eea
\bea
&&I(\tilde{\al},\tilde{\beta}) =
\frac{\Gamma(1+\lambda - \tilde{\al})}{
\Gamma(3-\tilde{\al}-\tilde{\beta})}
\frac{\pi \sin[\pi \tilde{\al}]}{\sin[\pi (\lambda-1+\tilde{\beta})\sin[\pi (\tilde{\al}+\tilde{\beta}+\lambda-1) ]}
- \frac{\Gamma(1+\lambda-\tilde{\al})}{\Gamma(3-\tilde{\al}-\tilde{\beta})}
\Psi(\lambda+\tilde{\al}-1)
\label{I4} \\
&&+ \sum_{n=0}^{\infty} \frac{\Gamma(n+2\lambda)}{n!(n+\lambda+1-\tilde{\al})} \biggl(\
\frac{\Gamma(n+2-\tilde{\al})\Gamma(2-\tilde{\beta})\Gamma(\lambda)
}{\Gamma(n+3+\lambda-\tilde{\al}-\tilde{\beta})\Gamma(3-\tilde{\al}-\tilde{\beta})\Gamma(\lambda+\tilde{\al}-1)}
\frac{\sin[\pi(\tilde{\beta}+\lambda-1)]}{\sin[\pi (\tilde{\al}+\tilde{\beta}+\lambda-1)]}
\nonumber \\
&& +  \frac{\Gamma(n+\tilde{\beta}+\lambda-1)}{\Gamma(n+2\lambda)} \Psi(n+2\lambda) \frac{\sin[\pi(\tilde{\beta}+\lambda-1)]}{\pi }
\biggl)
\, .
\nonum
\eea
%


\end{document}